\documentclass{article}
\usepackage{amsmath}
\usepackage{graphicx,psfrag,epsf}
\usepackage{enumerate}
\usepackage{natbib}
\usepackage{float}
\usepackage{verbatim}
\usepackage{url} 
\usepackage[normalem]{ulem}
\usepackage{comment}

\usepackage[table]{xcolor}
\newcommand{\blind}{0}

\usepackage{changepage}
\usepackage{soul}
\usepackage{xcolor}     
\usepackage[normalem]{ulem}

\usepackage{calc}
\newsavebox\CBox
\newcommand\hcancel[2][0.5pt]{%
  \ifmmode\sbox\CBox{$#2$}\else\sbox\CBox{#2}\fi%
  \makebox[0pt][l]{\usebox\CBox}%
  \rule[0.5\ht\CBox-#1/2]{\wd\CBox}{#1}}

\begin{document}

\bibliographystyle{agsm}

\def\spacingset#1{\renewcommand{\baselinestretch}%
{#1}\small\normalsize} \spacingset{1}


\if0\blind
{
  \title{\bf Integrating representative and non-representative survey data for efficient inference}
  \author{Nathaniel Dyrkton, Paul Gustafson, and Harlan Campbell\\
  Department of Statistics, University of British Columbia
  }
  \maketitle
} \fi

\if1\blind
{
  \bigskip
  \bigskip
  \bigskip
  \begin{center}
    {\LARGE\bf Title}
\end{center}
  \medskip
} \fi

\bigskip

\spacingset{1.45} 

\abstract{Non-representative surveys are commonly used and widely available but suffer from selection bias that generally cannot be entirely eliminated using weighting techniques. Instead, we propose a Bayesian method to synthesize longitudinal representative unbiased surveys with non-representative biased surveys by estimating the degree of selection bias over time. We show using a simulation study that synthesizing biased and unbiased surveys together out-performs using the unbiased surveys alone, even if the selection bias may evolve in a complex manner over time. Using COVID-19 vaccination data, we are able to synthesize two large sample biased surveys with an unbiased survey to reduce uncertainty in now-casting and inference estimates while simultaneously retaining the empirical credible interval coverage. Ultimately, we are able to conceptually obtain the properties of a large sample unbiased survey if the assumed unbiased survey, used to anchor the estimates, is unbiased for all time-points.}

\section{Introduction}

Representative survey methods rely on the conceptually attractive principle of probability-based sampling and have long been considered the gold standard \citep{lohr2019sampling}.  
Recent developments have, however, brought about a more nuanced understanding of the relative merits of representative and non-representative surveys. This shift is due to practical considerations (non-representative surveys are becoming increasingly fast and cheap, while representative surveys are becoming more difficult and expensive) as well as an evolving understanding of both the limitations and the potential of post-sampling adjustment methods.  
\vskip 0.2cm

It has long been recognized that obtaining unbiased estimates through post-sampling adjustment hinges on a crucial condition: systematic differences between the study sample and the target population must be known and measured, either directly or through proxies \citep{lohr2019sampling}.  However, the practical implications of this requirement remain somewhat unclear. Two recent papers, \citet{wang2015forecasting} and \citet{bradley2021unrepresentative}, underscore this ambiguity with their completely contradictory conclusions based on experiences with recent longitudinal surveys.
\vskip 0.2cm

\citet{wang2015forecasting} conclude that non-representative surveys can be used to generate accurate election forecasts by demonstrating that post-stratification and regression techniques can be used to correct for biases in a distinctly non-representative sample: voluntary participants solicited via the Xbox gaming platform.  The results of \citet{wang2015forecasting} suggest that post-sampling adjustment can correct for even extreme biases given that enough information about demographics, census, and historical data is available.  On the other hand, \citet{bradley2021unrepresentative} conclude that non-representative surveys (even when sample sizes are very very large) can be misleading after reviewing two large online surveys about COVID-19 vaccine uptake: (1) the Delphi-Facebook survey which recruited active Facebook users, and (2) the Census Household Pulse which, despite randomly sampling households (for which contact information was available), arguably failed to obtain a representative sample due to a very low response rate (5-7\%). These two surveys drastically overestimated the number of vaccinated Americans, even after careful post-sampling adjustment.  The  difference in \citet{wang2015forecasting} and \citet{bradley2021unrepresentative}'s conclusions is striking since both consider online surveys that use similar adjustment methods.  Are Xbox users somehow less prone to bias than Facebook users, or Census respondents? Are election forecasts easier to adjust for than vaccination rates?
\vskip 0.2cm

An intuitive explanation for the contradictory conclusions may be that \citet{wang2015forecasting} were able to adjust for a large number of key demographic variables while the two ``biased'' surveys that \cite{bradley2021unrepresentative} consider lacked certain key variables.  Indeed, neither the Delphi-Facebook nor Census Household Pulse surveys adjusted for the political partisanship of respondents, nor did they adjust for urbanicity (and the Delphi-Facebook survey did not explicitly adjust for education).  While these might seem like obvious and important omissions, selecting and accessing a sufficient set of variables on which to adjust is not trivial.  Indeed, in certain cases, perhaps counter-intuitively, adjusting for an additional seemingly important variable can increase rather than decrease bias. 
\vskip 0.2cm

Suppose, for example, that politically left-leaning individuals are more likely to be vaccinated, as are those living in urban areas.  Further, suppose that a survey over-represents those who are politically right-leaning and also over-represents those living in urban areas.  Ideally, one would collect information and adjust for both urbanicity and political leaning.  However, adjusting for neither might be better than adjusting for only one of the two.  Without any adjustment, the survey will be simultaneously biased downwards (estimating a lower vaccination rate) due to the excess of politically right-leaning respondents, and be biased upwards (estimating a higher vaccination rate) due to the excess of urban respondents. The biases might therefore cancel each other out, at least to some extent.  However, if one were to adjust for only political leaning and not for urbanicity, the biases would no longer cancel each other out, and the adjusted estimate would likely be more biased (upwards).
\vskip 0.2cm

If we draw our attention to the similarities between causal inference in observational studies and non-representative sampling, as suggested by \cite{mercer2017theory}, we can better understand the main assumptions required to obtain unbiased estimates. The key assumption for post-survey adjustments is that of ``conditional ignorability'' \citep{mercer2017theory,schuessler2023graphical}.  Much like the assumption of ``no unmeasured confounders''  that is required to obtain unbiased causal estimates in an observational study, the conditional ignorability assumption cannot be guaranteed.  In other words, there is no sure way of choosing a sufficient set of adjustment covariates (based on available data alone), and ultimately there is no test or way to validate when the assumption of conditional ignorability is met. 
\vskip 0.2cm





Does this mean non-representative surveys can never be entirely reliable and are of little value? Not necessarily.  When combined with representative surveys, non-representative surveys may still be valuable if they can be appropriately leveraged.
\vskip 0.2cm

Surprisingly, there is a lack of literature on combining non-representative surveys with representative surveys. One proposed method is weighting based on probability of inclusion conditioned on covariates \citep{elliot2009combining}. Another strategy is to weight using blended calibration, which combines unweighted non-probability surveys with weighted probability surveys by comparing to bench-marked values from the probability survey \citep{disogra2011calibrating}. \cite{wisniowski2020integrating} considers a Bayesian regression approach by combining small unbiased surveys with larger biased surveys, but find that bias can still be introduced by the inclusion of non-probability surveys. Notably, all of these proposed methods for combining non-probability surveys with probability surveys follow the same general principal of weighting based on auxiliary covariates, which may not leave the response given covariates conditionally ignorable on sample inclusion.
\vskip 0.2cm

In this paper, we propose a method that, instead of using weighting (of either individuals within surveys or of the entire surveys themselves), takes advantage of how bias changes over time within longitudinal surveys. Specifically, we develop a simple and fast Bayesian evidence synthesis method for combining non-representative longitudinal surveys with representative ones.  This method neither assumes previously applied weighting, nor requires any informative covariates.  One potential complication is that biases can change over time and we consider three possible ways to incorporate the evolution of bias starting from simplest to most complex.
\vskip 0.2cm

In Section 2, we outline the proposed method, state all assumptions, and provide an example where including biased surveys will reduce the uncertainty of estimates. In Section 3, we conduct a simulation study to show that using conservative assumptions, over-parametrizing the model is not overly detrimental, as long as the complexity of the model for the bias is proportional to the number of time-points.  In Section 4, inspired by the dataset presented by \cite{bradley2021unrepresentative}, we show that the proposed method performs well in estimating the vaccination rate by combining two ``biased'' surveys with a presumably ``unbiased'' survey. The results show a noticeable reduction in uncertainty (compared to only using the unbiased survey). We conclude in Section 5 by discussing potential uses and limitations for using large amounts of unrepresentative survey data to improve the precision of survey estimates.
\vskip 0.2cm
\section{Method}
Let us begin by describing the data and defining some basic notation.  Suppose we have data from $K$ surveys, each of which ask individuals a binary ``Yes or No'' question, and $N$ is the population size (assumed to be constant across time-points and surveys). Let $P_{t}$ be the total number of positive individuals (i.e., those who would, if asked, answer ``Yes'' to the survey question) at time-point $t$, where $t = 1$ is the first surveyed time-point. For group $k$ in $1, \ldots, K$, suppose:

\begin{itemize}
    \item $n_{kt}$ is sample size of the $k$-th survey at time-point $t$;
    \item $Y_{kt}$ is the number of positive (``Yes'') individuals amongst the $n_{kt}$ individuals surveyed in the $k$-th survey at time-point $t$.  We assume that individuals are surveyed such that they participate in only a single survey at only a single time-point.
\end{itemize}

 The goal is to estimate $\textrm{logit}^{-1}(\theta_{t})$, the  proportion of the (super-)population which is positive at time $t$, which is the expected value of  $P_{t}/N$. Specifically, one might be interested in one of three goals: general inference, ``now-casting'', and forecasting. General inference refers to collecting data up to time $T$ and making inferences on all time-points up to and including $T$ (i.e., inference on $\theta_0, \theta_1,..., \theta_T$). Now-casting refers to making inference on a specific time point using only the data collected prior to, and at, that time-point. Lastly, forecasting refers to making inferences about potential future parameters: $\theta_{T+1},\theta_{T+2},...$ . In this paper, we focus on general inference and now-casting.

 If the number of time-points is sufficiently large, a random walk model can be used to model how $\theta_t$ changes over time \citep{heidemanns2020updated}. For instance, suppose $P_t$ changes over time according to:
  \begin{eqnarray}
        \label{eq: P_t}
       P_{t} \sim \textrm{Binomial}(\textrm{logit}^{-1}(\theta_t), N),
 \end{eqnarray}
 where 
\begin{eqnarray}
        \label{eq: walk_theta}
       \theta_t|\theta_{t-1} \sim \textrm{Normal}(\theta_{t-1},\sigma^2),
 \end{eqnarray}
 and $\sigma^2$ represents the variance of the jump in the proportion of ``Yes" from the previous time-point (on the logit scale). 
 \vskip 0.2cm
 In non-representative surveys, estimating $\theta_t$ may be challenging due to selection bias. That is, individuals who would likely answer ``Yes" might be more inclined to participate in the survey than those who would likely answer ``No'' (or vice-versa). 
 \vskip 0.2cm
Let the degree of selection bias correspond to the $\phi_{kt}$ non-centrality parameter, where $Y_{kt}$ follows Fisher's non-central hyper-geometric distribution with
\begin{eqnarray}
    \label{eq: hypergeo}
(Y_{kt} |  P_{t}) \sim \textrm{NCHyperGeo}(P_{t}, N-P_{t}, n_{kt}, \phi_{kt}).
\end{eqnarray}
The hyper-geometric distribution describes the probability of drawing $Y_{kt}$ positive individuals amongst a sample of $n_{kt}$ who responded to the survey (without any individuals responding more than once), from a finite population of size $N$ that contains exactly $P_{t}$ positive individuals. The non-central hyper-geometric is a generalization of the hyper-geometric distribution whereby, in this case, the response may be biased so that either positive or negative individuals are more likely to respond.  When $\phi_{kt}>1$, positive individuals are more likely to respond than negative individuals;  when $\phi_{k}<1$, positive individuals are less likely to respond than negative individuals.  When $\phi_{k}=1$, we have that the probability of responding is equal for both positive and negative individuals, and the  non-central hyper-geometric distribution reduces to the standard hyper-geometric distribution. In this parameterization, the $\phi_{kt}$ parameter can be interpreted as an odds ratio: the odds of a positive individual being sampled vs. the odds of a non-positive individual being sampled. For a more detailed breakdown of the non-central hyper-geometric distribution, its use in population size estimation using surveys, and a discussion of Bayesian estimation of its parameters see \cite{ballerini2022fishers}.
\vskip 0.2cm
As discussed in the introduction, there may be several factors that lead to positive individuals being sampled, many of them may be completely unknown to the practitioner which ultimately makes survey weighting difficult. In this model, we assume there is a multitude of factors that contribute to selection bias and that we need not directly measure them. $\phi_{kt}$ is the resulting odds ratio of the unknown function of all of these factors.

In many cases, $N$ will be large (in relation to $n_{kt}$) and thus we may use an approximation to the non-central hyper-geometric distribution proposed by \cite{harkness1965properties}:
\begin{eqnarray}
    \label{eq: binom_approx}
     (Y_{kt}| \theta_t)  \sim \textrm{Binomial} \left( n_{kt}, \frac{p\phi_{kt}}{1-p + p\phi_{kt}}\right),
\end{eqnarray}
where $p = \textrm{logit}^{-1}(\theta_t)$.
This allows us to reduce potential computational difficulties associated with the probability mass function of the hyper-geometric distribution.
\vskip 0.2cm
Defining prior distributions is often controversial, as their choice can substantially influence the posterior when few data are available; see \citet{gelman2006prior}.  We proceed by adopting a truncated normal prior for $\sigma^2$:
\begin{flalign}
    \sigma^2 \sim \textrm{Normal}(0,\eta_0^2)\textrm{T}(0,\infty), \label{eq: sigmasqprior}
\end{flalign} 
where a larger $\eta^2_0$ expresses a greater prior uncertainty about $\sigma^2$. A suggested value is $\eta_0^2=1$, which corresponds to a prior belief that large changes in the positive rate, while possible, are rather unlikely. Specifically, if the positive rate at time $t-1$ were 0.5, at the next time point $\textrm{logit}^{-1}(\theta_{t})$ is between 0.26 and 0.76, 90\% of the time. For most purposes this is quite wide, unless the real-world time between $T$ and $T-1$ is in the vicinity of years. We may also adopt a normal prior on $\theta_0$:
\begin{equation}
    \theta_0 \sim \textrm{Normal}(\nu_0,\Gamma_0^2), \label{eq: theta0prior}
\end{equation}
where the value for $\nu_0$ directly encodes the belief of where the initial positive rate may lie and $\Gamma_0^2$ corresponds to the degree of certainty of this belief. For example, setting $\nu_0=0$ assumes the median of $\textrm{logit}^{-1}(\theta_0)$ is 0.5, which is likely appropriate for a tight presidential poll. Note that if the value for $\Gamma_0^2$ is too large, then the majority of the density of $\textrm{logit}^{-1}(\theta_0)$ will lie near the extremes (0 and 1). Thus a general prior we propose for most cases is $\nu_0 = 0$, and $\Gamma_0^2 = 2$,
where roughly 90\% of the mass of $\textrm{logit}^{-1}(\theta_0)$ lies between 0.14 and 0.86, and the density of $\textrm{logit}^{-1}(\theta_0)$ is approximately uniformly distributed, but its density decreases slightly near 0 and 1.
\vskip 0.2cm

 The only remaining component is a prior for $\phi_{kt}$.  The degree of selection bias 
 might vary considerably across surveys and may also vary, but perhaps to a lesser degree, across time. Given this parameter is latent and not expected to drastically change over time, we refrain from explicitly reviewing the interpretation of its prior. Instead, we propose basic wide priors that can be used for all but the most extreme purposes (such as the real-world time between time-points is years or decades). We consider three options for movement of $\phi_{kt}$ over time. Firstly, we consider that $\phi_{kt}$ is constant in time for each survey. Thus, for all $t$
 \begin{flalign} \label{eq: constphi}
     &\phi_{kt} = \exp(\gamma_{k}),  \\
     &\gamma_{k} \sim \textrm{Normal}(0,1). \nonumber
 \end{flalign}
Next we may consider that $\phi_{kt}$ increases or decreases slightly (and consistently) over time, thus $\phi_{kt}$ follows a linear model:
\begin{flalign}\label{eq: linearphi}
    &\phi_{kt}  = \exp(\gamma_{k0}+\gamma_{k1}t),\\ \nonumber
    &\gamma_{k0} \sim \textrm{Normal}(0, 1),\\ \nonumber
    &\gamma_{k1} \sim \textrm{Normal}(0, 0.25). \nonumber
\end{flalign}
Lastly, if $\phi_{kt}$ changes slightly (and more haphazardly) up or down relative to the previous survey collection:
\begin{flalign}\label{eq: walkphi}
      &\phi_{kt} =\exp(\gamma_{kt}),\\ \nonumber
      &\gamma_{kt}|\gamma_{k(t-1)} \sim \textrm{Normal}(\gamma_{k(t-1)},\pi^2),\\ \nonumber
      &\gamma_{k0}\sim \textrm{Normal}(0,1),   \\ \nonumber
       &\pi^2 \sim \textrm{Normal}(0,1)\text{T}(0,\infty), \nonumber
\end{flalign}
\noindent for the $k$-th survey at time-point $t$, for $k$ in 1,...,$K$.  
\noindent The prior specification therefore assumes that the degree of selection bias may be different for different studies and change over time but likely change slowly over time for a given study.
\vskip 0.2cm
In order to reasonably estimate $\phi_{kt}$ and the other model parameters, we need at least one unbiased survey to ``anchor'' the estimates. In this case we would have a subset of surveys for which $\phi_{kt}$ is known (usually equal to 1 for all $t$).  Without loss of generality, suppose this subset is the first $k^{'}$ surveys, such that for $k= 1, \ldots, k^{'}$, we have $\phi_{kt}=1$, for all $t$.  In a situation where all surveys are probability-based, $k^{'}=K$.

\begin{figure}[H]
    \centering
    \includegraphics[width = 14.5cm]{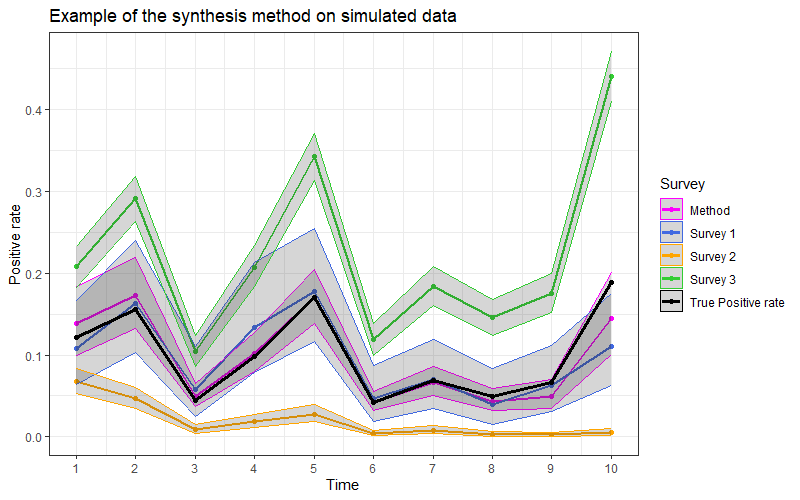}
        \caption{Example of the method on simulated data. The $\phi_{kt}$ assumption is linear  \eqref{eq: linearphi} where $k = 1$ is unbiased, and $k = 2$ \& $3$ correspond to biased surveys. Estimates are posterior medians and equal tailed 95\% credible intervals.}
    \label{fig:Proof of concept}
\end{figure}

To illustrate the proposed method, consider the goal of performing inference for $T = 10$ time-points and $K = 3$ surveys, with a population of $N = 10,000$. Survey 1 is unbiased, whereas survey 2 and 3 are compromised by selection bias.  Specifically, for these two surveys we generate data under the assumption of a linear $\phi_{kt}$  \eqref{eq: linearphi}. All the parameters and data are generated jointly using \eqref{eq: P_t},\eqref{eq: walk_theta} \eqref{eq: hypergeo}, and  \eqref{eq: sigmasqprior}, \eqref{eq: theta0prior}, \eqref{eq: linearphi} and the priors used in analysing the dataset are as such. The data are illustrated in Table \ref{tab:Table1_ProofOfConcept}.
\begin{table}[h]
    \centering  \renewcommand{\arraystretch}{1,3}
    \begin{tabular}{c|c|c|c|c|c|c|c|c|c|c}
                   \multicolumn{11}{c}{time-point}\\
               & 1 & 2 & 3 & 4 & 5 & 6 & 7 &8 & 9 & 10  \\
               \hline
      Survey 1 & $\frac{9}{100}$ & $\frac{18}{100}$ & $\frac{4}{100}$& $\frac{14}{100}$ & $\frac{20}{100}$ & $\frac{3}{100}$ & $\frac{8}{100}$ & $\frac{3}{100}$ & $\frac{6}{100}$ & $\frac{12}{100}$\\
      Survey 2 & $\frac{66}{1000}$ &  $\frac{48}{1000}$   & $\frac{7}{1000}$ & $\frac{19}{1000}$  & $\frac{30}{1000}$ & $\frac{2}{1000}$ & $\frac{10}{1000}$ & $\frac{2}{1000}$ 
 & $\frac{2}{1000}$ & $\frac{6}{1000}$\\
      Survey 3 & $\frac{207}{1000}$ &  $\frac{293}{1000}$   & $\frac{102}{1000}$ & $\frac{208}{1000}$  & $\frac{345}{1000}$ & $\frac{117}{1000}$ & $\frac{185}{1000}$ & $\frac{145}{1000}$ 
 & $\frac{174}{1000}$ & $\frac{441}{1000}$\\
     
    \end{tabular}
    \caption{Data generated for the illustrative example, each cell corresponds to $\frac{Y}{n}$ of the respective survey and time-point.}
    \label{tab:Table1_ProofOfConcept}
\end{table}
Analysis of each survey in isolation assumes $\phi_{kt} = 1$ for all $t$, but retain all the same priors \eqref{eq: P_t}, \eqref{eq: walk_theta}, \eqref{eq: hypergeo}, and \eqref{eq: sigmasqprior}, \eqref{eq: theta0prior}. In Figure \ref{fig:Proof of concept}, we see that using the biased surveys ($k = 2$ \& 3, assuming they are unbiased) in isolation, completely miss the true positive rate. On the other hand, the unbiased survey ($k = 1$) tracks the true positive rate well. The proposed method (magenta) also tracks the true positive rate, but the 95\% equal-tailed credible intervals more tightly encapsulate the true positive rate.
The reduction in credible interval width is quite noticeable. To quantify this, we take the mean ratio of the unbiased survey's credible interval width (in isolation) to that of the proposed method. Across all time-points this mean ratio is 2.18, meaning the method cuts the credible interval width in half. This reduction is present to a lesser degree in the now-casting case. The credible interval for the last time-point using the unbiased survey is only 1.12 times larger than that of the synthesis method. Moreover, the method's now-casted credible interval does contain the true positive rate whereas the unbiased survey's interval does not.
\section{Simulation study}
To assess the extent to which the synthesis method reduces the now-casted MSE and the effects of over-parametrizing the generating mechanism of $\phi_{kt}$, a 3 x 3 simulation design is performed. We consider 3 generating mechanisms for $\phi_{kt}$, where $\phi_{kt}$ either is constant  in time  \eqref{eq: constphi}, linear in time  \eqref{eq: linearphi}, or follows a random walk  \eqref{eq: walkphi}, all as described in the methods section. To get a good representation of how the methods perform under many parameter conditions we generate the parameters from the prior distributions. That is, each simulated dataset arises from different underlying parameter values. One issue from this technique is that priors presented in the methods section are wider than what might be realistic. For example, if one or more of the parameters are extreme in value, then $Y_{kt}$ may be equal to 0 or $n_{kt}$ for many values of $t$. To avoid this, we narrow the priors to more plausible values:
\begin{eqnarray*}
    \theta_0 \sim \textrm{Normal}(0,1),\\
    \sigma^2 \sim \textrm{Normal}(0,0.1)\textrm{T}(0,\infty),\\
    \gamma_1 \sim \textrm{Normal}(0,0.01),\\
    \pi^2 \sim \textrm{Normal}(0,0.01)\textrm{T}(0,\infty).
\end{eqnarray*}
This translates to $\textrm{logit}^{-1}(\theta_0)$ having roughly 90\% of its prior density between 0.22 and 0.78. Meaning most of the time we believe the initial positive rate is near 50\%. We also set narrow our belief about $\sigma^2$ which means
$\textrm{logit}^{-1}(\theta_t)-\textrm{logit}^{-1}(\theta_{t-1})$ is $\pm$0.15 or less, 90\% of the time, assuming $\textrm{logit}^{-1}(\theta_{t-1}) = 0.5$. For the linear model, $\phi_{kt}-\phi_{k(t-1)}$ in between -0.33 and 0.33  90\% of the time when $\phi_{k(t-1)} = 1$ . Lastly, for the random walk model $\phi_{kt}-\phi_{k(t-1)}$ is between -0.37 and 0.40, 90\% of the time, where $\phi_{k(t-1)}=1$.
\vskip 0.2cm
We simulate the data with $K=3$ surveys and with a population of $N=10,000,000$. Let $k = 1$ be the unbiased survey, and $k= 2$ and $3$ be the non-probability surveys. Given that non-probability surveys are often larger we set $n_{1t} = 100$, and $n_{2t} = n_{3t} = 1000$ for all $t$. We repeat this experiment for 3 sets of time-points: ($T$ = 5, 10, and 15). The models are all fit with 10 parallel chains using the dclone package \citep{solymos2010dclone} which links to  JAGS 4.3.x \citep{plummer2003jags}. Each chain has roughly 20,000-25,000 burn-in iterations and 50,000-70,000 draws depending on the complexity of $\phi_{kt}$, all with a thinning interval of 5.
\vskip 0.2cm
The goal is to compare the MSE, i.e., the averaged squared difference between $\textrm{logit}^{-1}(\theta_T)$ and the posterior median across the 2000 parameters generated. We add 95\% confidence intervals for the Monte Carlo MSE estimates based on the central limit theorem and the MCSE (Monte Carlo Standard Error), see \cite{morris2019}.

\subsection{Simulation considerations}
As mentioned before, the non-central hyper-geometric distribution \eqref{eq: hypergeo} introduces computational difficulties. Therefore the models are fit using the binomial approximation \eqref{eq: binom_approx}, but the data are still generated from \eqref{eq: hypergeo}, as we believe this is the true data generating mechanism. With $N = 10,000,000$ and the largest $n$ being $n_{kt} = 1000$, the binomial approximation is adequate. Random non-central hyper-geometric values are generated using the MCMCpack library \citep{martin2011mcmcpack}, which has an implementation based on work by \citet{liao2001fast}. Random values from the truncated normal are generated via the truncnorm library \citep{mersmann2018package}, and the rest of values are generated using R \citep{Rcore}. Moreover, for simulations with a large number of time-points, $\phi_{kt}$ can grow quickly for large values of $t$ (if $\phi_{kt}$ is linear,  \eqref{eq: linearphi})  and lead to computational issues. To fix this, the time-points, when the models are fitted, are centered in the $T = 10$ and $15$ simulations. Meaning: $\phi_{kt} = \exp(\gamma_{k0} + \gamma_{kt}t')$, where $t' =  t - \frac{T}{2}$. 
\subsection{Simulation Results}
\begin{figure}[H]
    \centering
    \includegraphics[width = 14cm]{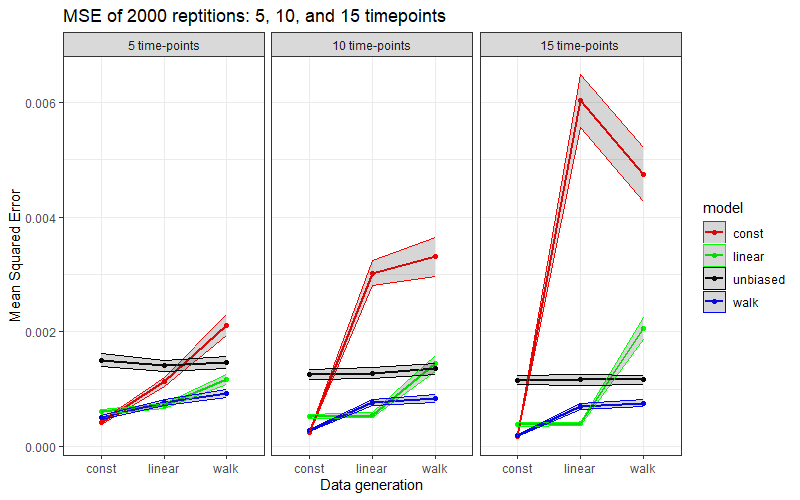}
    \caption{Results of simulation study for t = 5, 10, and 15 time-points. Simulations run using the model fit in JAGS 4.3.x. The MSE is calculated as the estimated averaged squared difference between $\textrm{logit}^{-1}(\theta_T)$ and the posterior median. Error bars represent the 95\% confidence intervals for the Monte Carlo estimated MSE i.e. $\widehat{\textrm{MSE}}\pm 1.96 \textrm{MCSE}$.
    }
    \label{fig:simulation_results}
\end{figure}
Figure \ref{fig:simulation_results} shows the results of the simulation study. As expected, when the model for $\phi_{kt}$ is correctly specified, the MSE is the lowest. Moreover, incorporating the additional 2 biased surveys generally leads to a lower MSE, showing the information from the biased surveys is useful. However, if the model for $\phi_{kt}$ is drastically over-parametrized, the synthesis model performs worse than using the unbiased survey in isolation. Specifically, fitting a constant $\phi_{kt}$, when $\phi_{kt}$ actually changes significantly, leads to a large increase in MSE. These simulations also suggest that the added complexity of fitting a random walk $\phi_{kt}$ does not lead to a large increase in estimator variance. Surprisingly, for scenarios with even a few time-points, a random walk model for $\phi_{kt}$ appears to be a relatively safe assumption. By increasing the number of time-points, additional complexity can be used to better adjust for the bias without gaining excessive estimator variance. Moreover, the confidence intervals of for Monte Carlo estimated MSE are sufficiently separated to have confidence in the results.
\section{Large survey vaccine data}
Monitoring the COVID-19 vaccination rates over time was an essential part of assessing the public health response to the pandemic. As discussed in the introduction, \cite{bradley2021unrepresentative} explains how two survey were found to drastically overestimate the true vaccination rate. Out of the three surveys analyzed, only the Axios-Ipsos was found to track the CDC's historically updated benchmark well (acknowledging imprecision in the benchmark). The Axios-Ipsos poll primarily focused on quality probabilistic sampling and maintained a high response rate (approx 50\%). As a result, its confidence intervals were found to contain the CDC's historically updated benchmark 10/11  times (up to June 2021), despite its small sample size. Thus, we propose using the method described in this paper to further improve the estimates of the Axios-Ipsos poll. We graciously use the extended data provided by \cite{bradley2021unrepresentative} to assess the method on real-world data. We aim to compare the performance of the synthesis method to the Axios-Ipsos Poll and the CDC's historically updated benchmark. We follow \cite{bradley2021unrepresentative} in adding $\pm 5\%$ error to the CDC's benchmark estimates.
Let $\phi_\textrm{AI} = 1$ for all $t$, and consider the three models for $\phi_\textrm{DF}$ and $\phi_\textrm{HP}$, where AI, DF, and HP refer to Axios-Ipsos, Delphi-Facebook, and Household-Pulse respectively. We consider two adjustments to the default prior settings suggested in Section 2.
Firstly, $\theta_0 \sim \textrm{Normal}(-2,1)$. Which means: roughly 90\% of the mass of $\textrm{logit}^{-1}(\theta_0)$ lies between 0.036 and 0.327. This reflects the prior belief in a low vaccination rate at time zero before seeing any data. Secondly, $\theta_t|\theta_{t-1} \sim \textrm{Normal}(\theta_{t-1},\sigma^2)\textrm{T}(\theta_{t-1},\infty)$ which represents the knowledge that the vaccination rate can only increase.
\subsection{Unequal Spacing}
The surveys are unequally spaced, that is, there is a survey measurement every week for the Delphi-Facebook survey, and much less frequent measurements for the other two surveys. This issue presents a challenge because a random walk depends on the most recent time-point. We select the dates of the Delphi-Facebook survey as the benchmark, and if any other survey has a measurement within the next 6 days of the benchmark, they are considered to have a measurement at the same time. Otherwise the number of vaccinated individuals for that survey are set as missing.
\subsection{Inference}
\begin{figure}[H]
    \centering
    \includegraphics[height = 7.5cm,width = 10cm]{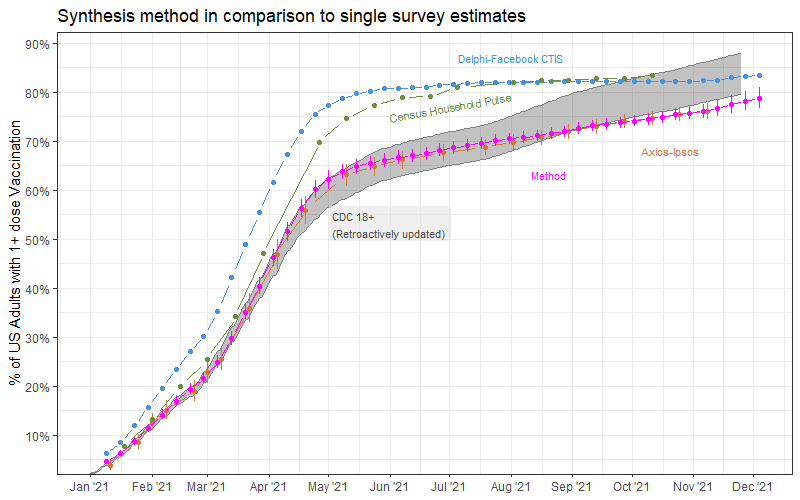}
    \caption{Plot of the result for assuming a random walk $\phi_{kt}$  \eqref{eq: walkphi}, with data extended from \cite{bradley2021unrepresentative}. Point estimates are posterior medians of the positive rate, and intervals are 95\% equal-tailed credible intervals. The CDC's historical benchmark has an assumed $5\%$ imprecision, see \cite{bradley2021unrepresentative} for details.}
    \label{fig:ExtendedWalk}
\end{figure}
Figure \ref{fig:ExtendedWalk} shows estimates obtained from applying the inference from our method in magenta, which shows a strong tendency to closely track the unbiased poll (Axios-Ipsos), with the added advantage of reducing the uncertainty in the estimation. 

\begin{figure}[H]
    \centering
    \includegraphics[height = 7.5cm,width = 10cm]{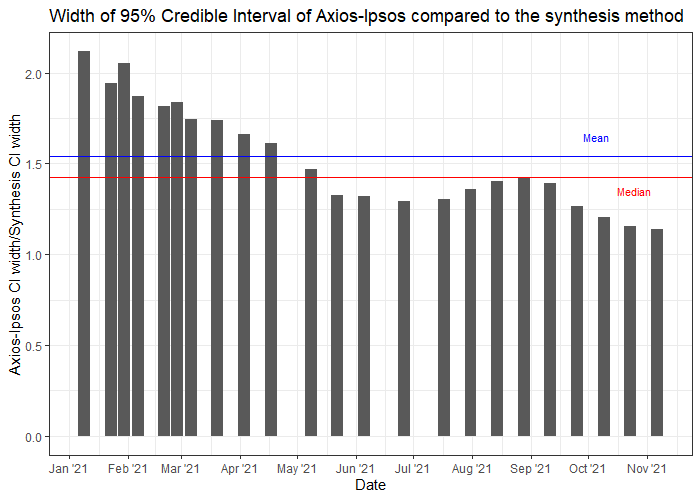}
    \caption{Ratio of 95\% equal-tailed credible interval widths of Axios-Ipsos surveys and the synthesis method.}
    \label{fig:ReductionInInferenceWidth}
\end{figure}
In Figure \ref{fig:ReductionInInferenceWidth} we see that the Axios-Ipsos survey alone has a larger 95\% credible intervals for all time-points. The average and median Axios-Ipsos credible interval widths are 1.54 and 1.42 times larger respectively (where the Axios-Ipsos survey is not missing).
This demonstrates the clear advantage of synthesising the information from multiple surveys, even if we acknowledge that the bias may be changing over time. The unbiasedness property of the Axios-Ipsos survey also appears to be preserved. The synthesis method's 95\% credible intervals are within the CDC's historically updated benchmark (including the assumed CDC's 5\% margin of error) $43/46$ (93.5\%) of the time (this excludes the last two time-points where no CDC data is present). 
\begin{figure}[H]
    \centering
    \includegraphics[height = 5.1cm,width = 7cm]{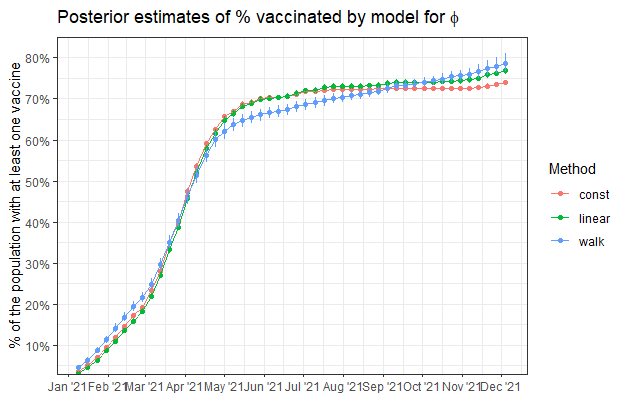}
    \includegraphics[height = 5.1cm,width = 7cm]{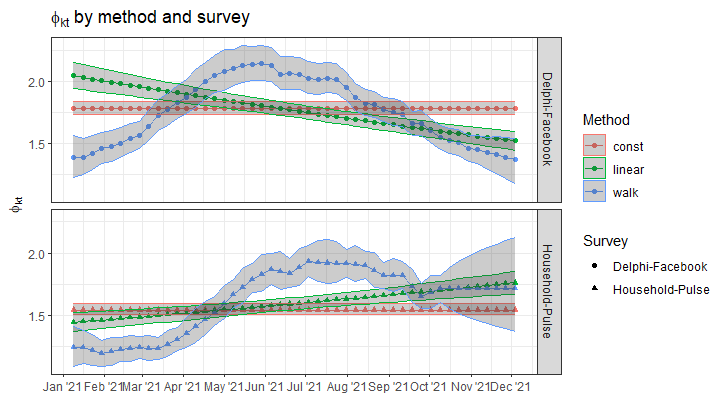}
    \caption{Plot of the estimates of vaccination rate by assumption on $\phi_{kt}$. Right-side shows the estimates of $\phi_{kt}$ for the biased surveys by method (posterior medians + equal tailed 95\% credible intervals).}
    \label{fig: comparisons}
\end{figure}
Figure \ref{fig: comparisons} shows that changing the specification for $\phi_{kt}$ changes the estimates of the model significantly, as near the tail end of dates, posterior medians of the vaccination rate can differ by around 10\%. Moreover, if the model for $\phi_{kt}$ is specified as constant, this model predicts a drastically different curve, especially for the later time points. This aligns with the findings of \cite{bradley2021unrepresentative}, in which the estimated bias is shown to change drastically over time. The MCMC chains for the constant $\phi_k$ model also fails to converge, with $\hat{R} > 1.1$ for all parameters, which may give unreliable estimates. This is more evidently seen with the plot on the right: the random walk $\phi_{kt}$ moves both up and down, suggesting neither a linear nor constant $\phi_{kt}$ is an appropriate simplification. We can assess the validity of the random-walk model by comparing the posterior estimates of $\phi_{kt}$ to the bias estimated by \cite{bradley2021unrepresentative}. In Figure 1b of \cite{bradley2021unrepresentative}, they show the total error by surveys: the survey estimate minus the truth. These estimates have a similar direction as posterior bias estimates presented (up to time-point 20 or June 2021). However, the estimates presented here are not entirely comparable up to June 2021 because information from later time-points affects the estimates of the earlier time-points.

\subsection{Now-cast performance}
As well as making inference for each time-point after collecting the data on all time-points we also consider the now-casting performance. That is we only consider the information up to and including each time-point. In this section, we compare the now-cast point estimate and 95\% credible interval for each of the 48 time-points. Each of the Axios-Ipsos, Household-Pulse, and Delphi-Facebook surveys are fit individually using the specified models in the methods section with $\phi_{\cdot t} = 1$, for all $t$, and our method is fit using information from the three surveys up to and including each point with a random walk assumption for $\phi_{DF}$ and $\phi_{HP}$.
\begin{figure}[H]
    \centering
    \includegraphics[height = 7.5cm,width = 10cm]{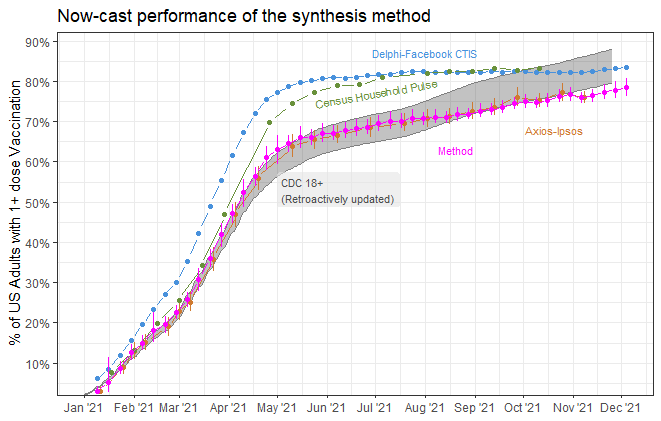}
    \caption{Plot of now cast performance of synthesis method. Estimates are posterior medians and equal-tailed 95\% credible intervals. Figure extends data from \cite{bradley2021unrepresentative}.}
    \label{fig:WalkPhi}
\end{figure}
In Figure \ref{fig:WalkPhi} we can see the point estimates are similar the inference case (Figure \ref{fig:ExtendedWalk}), but with a considerable amount of added uncertainty. Again, the uncertainty of the synthesis method is much lower than that of the Axios-Ipsos survey used in isolation. Precisely, over the time-points where the Axios-Ipsos poll is not missing, the mean and median ratio of the Axios-Ipsos credible interval widths to that of the proposed method are both 1.24 times larger. This decrease in uncertainty does not appear to reduce the empirical credible interval coverage. The synthesis method's 95\% credible intervals intersect with the CDC's historically updated benchmark range $44/46$ (95.7\%) of the time. Moreover, there are a few time-points at which the method yields extremely large credible intervals, such as time-points 2 and 6. These are a result of the missing values in the unbiased data. Especially at the beginning of the time-points when few Axios-Ipsos surveys are seen, a missing value leads to a relatively wide estimation of $P(Y_{kt}|.)$, which ultimately increases the uncertainty about
$\textrm{logit}^{-1}(\theta_t)$.
\begin{figure}[H]
    \centering
    \includegraphics[height = 6cm,width = 7cm]{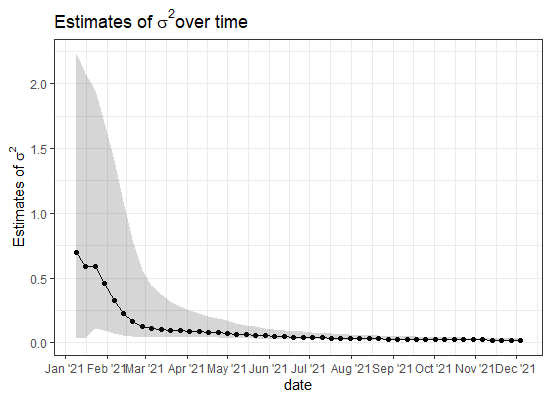}
    \includegraphics[height = 6cm,width = 7cm]{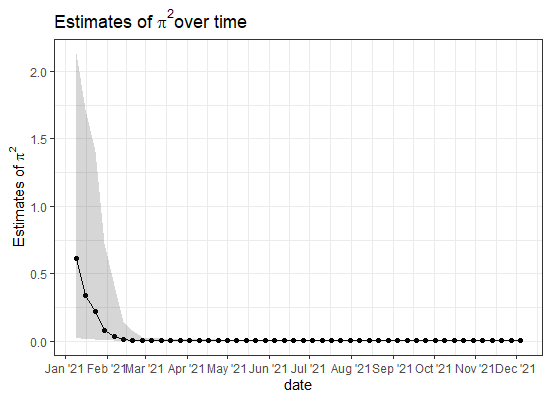}
    \caption{Now-casted estimates (posterior medians + equal-tailed 95\% credible intervals) for $\sigma^2$ and $\pi^2$, the jumping variance of the positive rate and variance of $\gamma$.}
    \label{fig: Sigmasq and Pi}
\end{figure}
Figure \ref{fig: Sigmasq and Pi} shows the posterior median and 95\% credible intervals for $\sigma^2$ and $\pi^2$ across all 48 time-points. Note that for time-point 1, the estimate of $\sigma^2$ is essentially equal to that given by the prior. Around mid September there is a slight drop in the estimate for $\sigma^2$. This is because as the number of individuals who have been vaccinated starts to level off for both the Delphi-Facebook and Household-Pulse surveys. The uncertainty around $\pi^2$ also reduces quite quickly. After only about 7-8 time-points most of belief of the variance in the movement in $\phi_{kt}$ has been narrowed.

\subsection{Inclusion of biased surveys quantified in terms of increased unbiased sample size}
One way to measure the reduction in uncertainty is to measure increase in iid sample size in the unbiased sample required to obtain the same level of uncertainty as the synthesis method. To measure this, we fit the method with the only the Axios-Ipsos data, and then compare the width of the credible intervals of the Axios-Ipsos survey against that of combining the Axios-Ipsos and one of the other surveys, and then ultimately both surveys. The comparison is only made for time-points where an Axios-Ipsos survey is taken. We proceed by calculating $n$ based on the classical frequentist confidence intervals.
\begin{itemize}
    \item Let $\hat{p}_t$ be the positive rate estimate for time point $t$;
    \item Let $R_t$ be the ratio of credible interval widths of the method to the baseline Axios-Ipsos poll;
    \item Let $\text{MOE}_t$ be the margin of error of the credible interval at time t.
\end{itemize}
Then we can easily get the number of $n_\textrm{iid}$ required for a $(1-\alpha/2)100\%$ confidence interval:
\begin{eqnarray}
    n_{\text{iid } t} = \frac{Z_{1-\alpha/2}^2 \hat{p}_t(1-\hat{p}_t)}{ R_t \text{MOE}_t}.
\end{eqnarray}
\begin{figure}[H]
    \centering
    \includegraphics[height = 7.5cm,width = 12cm]{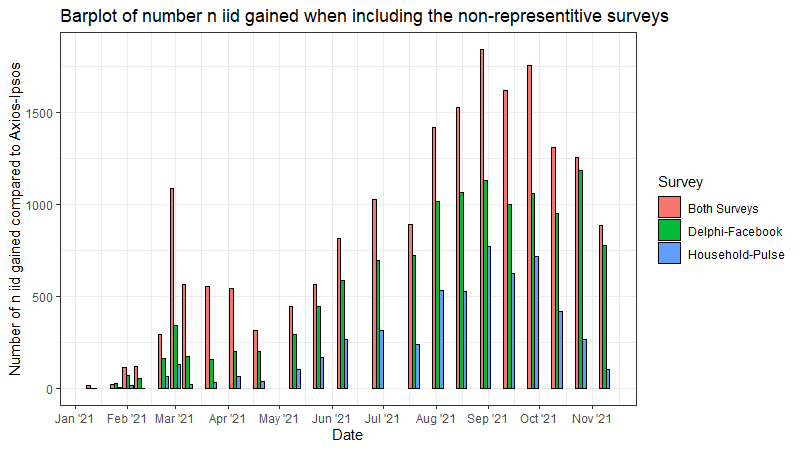}
    \caption{Plot of number of the number of $n_\textrm{iid}$ gained by date, for each survey. The bars represent $n_\text{iid Synthesis}-n_\text{iid Axios-Ipsos}$ at the 95\% confidence level.}
    \label{fig:n_gained}
\end{figure}
\begin{table}[H]
    \centering
    \begin{tabular}{c|c|c}
    \hline
            Surveys included        & mean gain & median gain \\
                    \hline \hline
      All surveys   &  825 & 817 \\
      Delphi-Facebook  & 534 & 443\\
      Household-Pulse & 235 & 131\\
      \hline
    \end{tabular}
    \caption{Summary statistics on the effective number of $n_\textrm{iid}$ gained from the inclusion of the biased surveys, rounded to the nearest integer.}
    \label{tab: n_gained}
\end{table}
Figure \ref{fig:n_gained} shows the plots of the gain in $n_\textrm{iid}$ by time-point. We can see that there is considerable information gained for later time-points. The first and fourth time-point have small negative gains when the Household-Pulse survey is included. This is likely due to there being two missing values in four time-points for the Household-Pulse survey (and one missing Axios-Ipsos time-point in between) leading to large posterior variances for parameters. Thus many of these bars are not completely accurate given many of the surveys are missing for some time-points, and we would expect the gain to be larger if all three surveys were present for all time-points. Ultimately, synthesizing a larger survey (in terms of sample size) translates into a larger gain in $n_\textrm{iid}$ for the synthesis method. Meaning we could conceptually achieve the properties of an unbiased large sample size survey by combing a small unbiased survey with an extremely large biased survey.

\section{Discussion}
\subsection{Limitations and unbiased anchor surveys}
We have demonstrated that it is not only possible to synthesize many potentially biased surveys with an unbiased representative survey, but it is advantageous to do so. In aggregate, the proposed synthesis method closely tracks the unbiased survey and reduces posterior variance. However, the proposed method relies on the rather strong assumption that the unbiased anchor survey is truly unbiased (or if the bias is completely known for all $t$). In reality, this assumption can be either difficult or impossible to meet. It may be feasible if the anchor survey overwhelmingly focuses on the quality of sampling design to achieve properties of a probability sample, rather than simply focusing on sample size. If this is the case, we can achieve the properties of a large sample unbiased survey by synthesizing a small unbiased survey with one or more large online surveys, which are easily available. In the event that an unbiased survey is not present, we may use a well-weighted and carefully designed survey as the ``unbiased" or anchor survey. However, as discussed in the introduction, we may again run into the issue of being unable to meet conditional ignorability. Picking a well weighted anchor survey can still provide substantial benefits by reducing posterior uncertainty about the positive rate if the belief in the weighted survey is justified. Admittedly, if the reference or anchor survey is not properly weighted, using the provided method could worsen survey estimates as we may increase bias significantly, relative to the true positive rate, with a reduction of uncertainty around said biased estimate. This can be easily understood if in the large survey vaccination example we chose the Delphi-Facebook survey as the anchor survey. Therefore, the weighting procedure of a the anchor survey should be rigorous, and other factors such as response rate and reputation of the survey practitioner should be considered.

\vskip 0.2cm

\subsection{Selecting a model for $\phi$}
In this paper we have provided three possible models for how $\phi$ may evolve over time: the constant model, the linear model, and the random walk model. The choice of selecting the model for $\phi$ depends on the prior belief about the evolution of the bias and the bias variance trade-off of over-parametrizing or under-parameterizing the model. In most applications the bias would neither be constant nor linear, but interpolating the crude error of a random walk with a linear model for $\phi$ when there are few time-points is a reasonable choice. However, as shown in Figure \ref{fig:simulation_results}, choosing a random walk model is a conservative and surprisingly efficient choice. There is also the possibility of extending the complexity of the model for $\phi$ if the number of time-points is large enough. We could consider employing a moving average model or adding other higher order auto-regressive terms. Yet, we must also be careful about adding too much complexity to a latent parameter, but the possibility exists.
Furthermore, we have only considered fitting the same model for both of the non-representative surveys. There is an argument to be made to either specify different models for $\phi_k$ for each $k$, or to force the non-representative surveys to share the same jump or slope. In the latter case, this assumes the bias $\phi$ is not necessarily a parameter of the survey, but a parameter of the population that surveys sample from. In either case, the most conservative model (random walk) shown in this paper can be relied upon, having been shown to have a lower MSE than using the unbiased survey, even when over-parametrized.

\subsection{Problems with identifying $\phi$}
We briefly mentioned in the simulation section that one can run into an issue if $Y_{kt} = n_{kt}$ for at least one $k$ or $t$ in the biased surveys. Suppose that $Y_{kt} = n_{kt}$ for at least one $k$, $t$ in the biased surveys, and  $Y_{kt}\leq n_{kt}$ for the unbiased survey(s). Consequently, $\phi_{kt}$ may lie anywhere in $[1,\infty)$, and so the posterior estimate of $\phi_{kt}$ would revert to something resembling the prior. The non-central hyper-geometric model \eqref{eq: hypergeo} cannot be used if $\phi_{kt}$ is unidentifiable for any $k,t$, as JAGS will not run. It is not completely understood why JAGS cannot handle these cases, as the binomial approximation \eqref{eq: binom_approx} retains the odds ratio interpretation and runs as expected with reasonable results. This discrepancy is likely related to the complex probability mass function of the non-central hyper-geometric distribution. The question remains: does the unidentifiable $\phi$ produce excessive problems in inference or now-casting? By exclusively analyzing the simulation results where $\phi$ is unidentifiable, there is no obvious change in the bias, and the large reduction in variance is still present. Yet this may be due to the correct specification of the priors. In practice, we would hope that there is a rarely a case where surveys have such a large selection bias where the values of $Y_{kt}$ are so extreme.
\subsection{Extensions to more general data}
The proposed method in this paper is only suited for a binary response. However, extensions to a continuous response would be straightforward. The key idea is that we track how the survey's response bias changes over time. We need not restrict ourselves to any specific type  response type, but instead rely on the central limit theorem. For instance, suppose we are interested in surveying salaries and it is presumed that individuals may over-represent their salaries. In this case, a given survey's estimate may be log normal. Let a given survey $k$ estimate at time $t$ be $\bar{x}_{kt}$, then the sampling distribution of survey estimates could be $(\bar{X}_{kt}|\mu_{t},\sigma^2,\phi_{kt}) \sim N(\mu_{t} + \phi_{kt},\sigma^2(1-\frac{n_{kt}}{N}))$. Here $\phi_{kt}$ no longer represents an odds ratio, but rather an additive bias that a survey receives from selection bias and/or biased responses. 
\subsection{The take-away}
Ultimately, we provide a framework for combining non-representative and representative surveys to reduce uncertainty in survey estimates provided an unbiased survey is present. We can achieve the desirable properties of an unbiased sample, while simultaneously taking advantage of the numerous online surveys that have a large sample size, but suffer from selection bias. These conceptual properties are shown to hold if the movement of selection bias over time is complex. We've also provided a model for $\phi_{kt}$ that reliably reduces uncertainty in estimates, even when over-paramterizing. The framework can be extended to incorporate multiple types of data, and it makes no assumptions on weighting procedures which leaves the door open to practitioners who prefer to weight before making inferences.

\bibliography{truthinscience}

\end{document}